# The Equivalence of Semidefinite Relaxation MIMO Detectors for Higher-Order QAM

W.-K. Ma, C.-C. Su, J. Jaldén, T.-H. Chang, and C.-Y. Chi

**Abstract**

In multi-input-multi-output (MIMO) detection, semidefinite relaxation (SDR) has been shown to be an efficient high-performance approach. Developed initially for BPSK and QPSK, SDR has been found to be capable of providing near-optimal performance (for those constellations). This has stimulated a number of recent research endeavors that aim to apply SDR to the high-order QAM cases. These independently developed SDRs are different in concept and structure, and presently no serious analysis has been given to compare these methods. This paper analyzes the relationship of three such SDR methods, namely the polynomial-inspired SDR (PI-SDR) by Wiesel *et al.*, the bound-constrained SDR (BC-SDR) by Sidiropoulos and Luo, and the virtually-antipodal SDR (VA-SDR) by Mao *et al.* The result that we have proven is somehow unexpected: the three SDRs are equivalent. Simply speaking, we show that solving any one SDR is equivalent to solving the other SDRs. This paper also discusses some implications arising from the SDR equivalence, and provides simulation results to verify our theoretical findings.

This work was supported in part by National Science Council, R.O.C., under Grants 96-2628-E-007-002-MY2 and NSC 96-2219-E-007-001; and by a General Research Fund of Research Grant Council, Hong Kong. Dr. Jaldén was supported by the STREP project No. IST-026905 (MASCOT) within the Sixth Framework Programme of the European Commission. Part of this work was presented in ICASSP2008.

Wing-Kin Ma is with Department of Electronic Engineering, The Chinese University of Hong Kong, Shatin, N.T., Hong Kong. E-mail: wkma@ieee.org

Chao-Cheng Su, Tsung-Hui Chang, and Chong-Yung Chi are with Institute of Communications Engineering & Department of Electrical Engineering, National Tsing Hua University, Hsinchu, Taiwan, R.O.C. E-mails: g9564512@oz.nthu.edu.tw, d915691@oz.nthu.edu.tw, cychi@ee.nthu.edu.tw

Joakim Jaldén is with Institute of Communications and Radio-Frequency Engineering, Vienna University of Technology, Austria. E-mail: joakim.jalden@nt.tuwien.ac.at

I. INTRODUCTION

Multiple-input-multiple-output (MIMO) detection using semidefinite relaxation (SDR) [1]–[14] has recently received increasing attention. Being a symbol-constellation dependent technique, SDR has been shown to provide considerably better symbol error performance than some suboptimal MIMO detectors such as the linear and decision-feedback receivers. SDR is not an optimal approach from a maximum-likelihood (ML) perspective, but it guarantees a polynomial-time complexity in the problem dimension. In comparison, the currently best known methods for optimal ML MIMO detection, namely sphere decoding [15], [16] do not have such a guarantee [17].[1]

SDR was first proposed for the BPSK constellation [1], [2], and the same idea can easily be carried forward to the QPSK constellation (or 4-QAM) [6], [7]. For BPSK and QPSK, simulation results have indicated that SDR can provide near-optimal bit error performance. This intriguing finding has stimulated a number of works. Theoretically, it is shown recently [13] that SDR can achieve the full receive diversity in the BPSK case. An equally interesting but totally different analysis is given in [14], where the SDR approximation gap is examined using random matrix theory. Apart from theoretical analysis, there has been interest in various aspects such as fast implementations [19], [20] and soft-in-soft-out MIMO detection [7].

But what attracts more attention in SDR is possibly its extension to more general symbol constellations, especially the higher-order QAM. For higher-order PSK, readers are referred to [3]. For higher-order QAM which is the focus of this paper, the first endeavor is by Wiesel *et al.* [9], who proposed a polynomial-inspired SDR (PI-SDR) method for 16-QAM. In that work the authors also showed that PI-SDR is a bidual of the optimal ML MIMO detector (or achieves an optimal Lagrangian dual lower bound of the ML). The drawback of PI-SDR is that its extension to larger QAM sizes would be increasingly complex to handle. Later, Sidiropoulos and Luo proposed a bound-constrained SDR (BC-SDR) method [9] for any QAM. BC-SDR aims at simplicity and appears to be less sophisticated than PI-SDR. For instance, the BC-SDR problem structure is virtually the same for any QAM size. The simple, special structures of BC-SDR make fast implementations [21] particularly favorable. More recently, Mao *et al.* [10] developed a virtually-antipodal SDR (VA-SDR) method for any $4^q$-QAM (where $q \geq 1$). As its name implies, VA-SDR has a strong connection to the SDR used in BPSK/QPSK. VA-SDR is structurally less complex than PI-SDR, but involves more optimization variables than BC-SDR.

---

[1]As a short aside, the complexity limitation of optimal sphere decoding has recently motivated interest in some suboptimal but reduced-complexity variants; e.g., the Fano decoder [16] and the fixed-complexity sphere decoder [18].

It is worthwhile to mention two more recent developments. Mobasher *et al.* formulated a class of SDR problems that is applicable to any kind of symbol constellations [12]. As a price for their generality, Mobasher's formulations are considerably more complex than the other SDRs. This translates into a higher computational requirement. Yang *et al.* [11] proposed a tightened version of BC-SDR for the 16-QAM case. Interestingly, they showed that their tightened BC-SDR can provide a better approximation than the 16-QAM PI-SDR.

While a number of SDR methods have been proposed for higher-order QAM, their comparisons have not been seriously considered at present. This paper focuses on analyzing the relationship of the PI-SDR, BC-SDR, and VA-SDR methods. We obtain a result that is intuitively not so obvious: *PI-SDR, BC-SDR, and VA-SDR are actually equivalent, despite the fact that they exhibit rather different problem structures.* Specifically, our analysis reveals that

1) For 16-QAM and 64-QAM, PI-SDR is equivalent to BC-SDR.
2) For any $4^q$-QAM, VA-SDR is equivalent to BC-SDR.

The SDR equivalence result to be established is quite general, and it can be applied to problems other than MIMO detection. For example, the SDR equivalence is perfectly applicable to the problem of blind ML detection of orthogonal space-time block codes [22] where the objective function structure is different from that in MIMO detection.

We attempt to give insights into the SDR equivalence result by organizing this paper in the following way. We use the relatively simple case of 16-QAM to provide the problem statement in Section II, and to review the three SDR methods in Section III. This is followed by Section IV, where we describe the 16-QAM SDR equivalence, illustrate the ideas behind, and discuss its implications. Then, Section V addresses the more complex case where the QAM size is further increased. Some simulation results are provided in Section VI for verifying our claim and for demonstrating the SDR performance.

## II. PROBLEM STATEMENT

We consider a standard MIMO system model, namely that of spatial multiplexing (or V-BLAST) over a frequency-flat channel [23], [24]. Supposing that the system has $\tilde{N}$ transmitter antennas and $\tilde{M}$ receiver antennas, the received signal can be described by the following formula:

$$\tilde{\mathbf{y}} = \tilde{\mathbf{H}}\tilde{\mathbf{s}} + \tilde{\boldsymbol{\nu}}. \tag{1}$$

Here, $\tilde{\mathbf{H}} \in \mathbb{C}^{\tilde{M} \times \tilde{N}}$ is the MIMO channel, $\tilde{\mathbf{y}} \in \mathbb{C}^{\tilde{M}}$ is the received signal vector, $\tilde{\boldsymbol{\nu}}$ is a noise vector assumed to be zero-mean circular white Gaussian, and $\tilde{\mathbf{s}} \in \mathcal{S}^{\tilde{N}}$ is the transmitted symbol vector where $\mathcal{S} \subset \mathbb{C}$ denotes the symbol constellation set. For the 16-QAM constellation, $\mathcal{S}$ is given by

$$\mathcal{S} = \{\, s = s_R + js_I \mid s_R, s_I \in \{\pm 1, \pm 3\} \,\}.$$

It should be stressed that in many digital communication problems, the signal models can be formulated in the same form as in (1); for example, the CDMA multiuser problem. Readers can find more examples in the literature such as [15], [16]. Thus, detection techniques for (1), or simply MIMO detection, have wide applicability.

It would be convenient to reformulate the complex-valued model in (1) to a real-valued one. Let

$$\mathbf{y} = \begin{bmatrix} \Re\{\tilde{\mathbf{y}}\} \\ \Im\{\tilde{\mathbf{y}}\} \end{bmatrix}, \quad \mathbf{s} = \begin{bmatrix} \Re\{\tilde{\mathbf{s}}\} \\ \Im\{\tilde{\mathbf{s}}\} \end{bmatrix}, \quad \mathbf{H} = \begin{bmatrix} \Re\{\tilde{\mathbf{H}}\} & -\Im\{\tilde{\mathbf{H}}\} \\ \Im\{\tilde{\mathbf{H}}\} & \Re\{\tilde{\mathbf{H}}\} \end{bmatrix},$$

$M = 2\tilde{M}$, and $N = 2\tilde{N}$. Eq. (1) is equivalent to

$$\mathbf{y} = \mathbf{H}\mathbf{s} + \boldsymbol{\nu} \qquad (2)$$

where $\mathbf{s} \in \{\pm 1, \pm 3\}^N$ and $\boldsymbol{\nu}$ is defined in the same way as $\mathbf{y}$. The ML detection problem for the MIMO model in (2) is shown to be an optimization

$$\min_{\mathbf{s} \in \{\pm 1, \pm 3\}^N} \|\mathbf{y} - \mathbf{H}\mathbf{s}\|^2, \qquad (3)$$

in which the globally optimal solution serves as the ML decision. Note that $\|\cdot\|$ in (3) stands for the vector 2-norm. ML detection is known to provide superior detection performance, but the major challenge lies in solving (3) which is a computationally hard problem. Presently, the best known optimal solver for (3) is sphere decoding [15], [16]. While sphere decoding has been empirically found to be computationally very fast for small to moderate problem sizes (say, for $N \leq 20$ for 16-QAM), it has been revealed [17] that the sphere decoding complexity would be prohibitive for large $N$ and/or low SNRs.

## III. Review of Three 16-QAM SDR Detectors

SDR is a suboptimal approach to ML, using a class of polynomial-time solvable convex optimization problems known as semidefinite programs. In this section, we review three SDR methods for the 16-QAM constellation, namely PI-SDR [8], BC-SDR [9], and VA-SDR [10]. (Their extensions beyond 16-QAM will be considered later in the paper.)

*A. Polynomial Inspired SDR*

PI-SDR was the first application of the SDR principle [25] to 16-QAM ML detection, to the best of our knowledge. To present its idea, consider a reformulation of the ML problem in (3)

$$\min_{\mathbf{S} \in \mathbb{S}^N, \mathbf{s} \in \mathbb{R}^N} \quad \operatorname{tr}(\mathbf{H}^T \mathbf{H} \mathbf{S}) - 2\mathbf{s}^T \mathbf{H}^T \mathbf{y} + \|\mathbf{y}\|^2 \qquad (4)$$
$$\text{s.t.} \quad \mathbf{S} = \mathbf{s}\mathbf{s}^T, \quad S_{ii} \in \{1, 9\}, \quad i = 1, \ldots, N$$

where $\mathbf{S}$ is a slack variable, $\mathbb{S}^N$ is the set of $N \times N$ real symmetric matrices, $S_{ij}$ denotes the $(i,j)$th element of $\mathbf{S}$, and $\operatorname{tr}(\cdot)$ is the trace operator. PI-SDR was inspired by the fact that

$$u \in \{1, 9\} \iff (u-1)(u-9) = 0 \iff u^2 - 10u + 9 = 0.$$

By turning the constraints $S_{ii} \in \{1, 9\}$ to a polynomial form, Problem (4) is further reformulated as

$$\min_{\mathbf{S}, \mathbf{s}, \mathbf{U}, \mathbf{u}} \quad \operatorname{tr}(\mathbf{H}^T \mathbf{H} \mathbf{S}) - 2\mathbf{s}^T \mathbf{H}^T \mathbf{y} + \|\mathbf{y}\|^2$$
$$\text{s.t.} \quad \mathbf{S} = \mathbf{s}\mathbf{s}^T, \quad \mathbf{U} = \mathbf{u}\mathbf{u}^T \qquad (5)$$
$$d(\mathbf{S}) = \mathbf{u}, \quad d(\mathbf{U}) - 10\mathbf{u} + 9\mathbf{1}_N = \mathbf{0}$$

where $d : \mathbb{R}^{N \times N} \to \mathbb{R}^N$ is the diagonal operator (i.e., $d(\mathbf{S}) = [\ S_{11}, \ldots, S_{NN}\ ]^T$), and $\mathbf{1}_N$ is the $N$-dimensional all-one vector. The reformulated ML problem in (5) is still hard, where the difficulty lies in the nonconvex constraints $\mathbf{S} = \mathbf{s}\mathbf{s}^T$ and $\mathbf{U} = \mathbf{u}\mathbf{u}^T$ which restrict $\mathbf{S}$ and $\mathbf{U}$ to be of rank 1.

In PI-SDR, we relax the polynomial ML formulation in (5) to

$$\min \quad \operatorname{tr}(\mathbf{H}^T \mathbf{H} \mathbf{S}) - 2\mathbf{s}^T \mathbf{H}^T \mathbf{y} + \|\mathbf{y}\|^2$$
$$\text{s.t.} \quad \mathbf{S} \succeq \mathbf{s}\mathbf{s}^T, \quad \mathbf{U} \succeq \mathbf{u}\mathbf{u}^T \qquad (6)$$
$$d(\mathbf{S}) = \mathbf{u}, \quad d(\mathbf{U}) - 10\mathbf{u} + 9\mathbf{1}_N = \mathbf{0}$$

where $\mathbf{A} \succeq \mathbf{B}$ means that $\mathbf{A} - \mathbf{B}$ is positive semidefinite (PSD). The idea is to replace the hard constraint $\mathbf{S} = \mathbf{s}\mathbf{s}^T$ by a convex constraint $\mathbf{S} \succeq \mathbf{s}\mathbf{s}^T$, and similarly to $\mathbf{U} = \mathbf{u}\mathbf{u}^T$. There are two basic advantages with such a relaxation. First, Problem (6), or the PI-SDR problem is a semidefinite program (SDP) which is convex and does not suffer from local minima. Second, as an SDP the PI-SDR problem can be solved by available interior-point methods [26], [27] in a polynomial-time fashion.

Once we solve the PI-SDR problem in (6), we can make a symbol decision by simple rounding of the PI-SDR solution associated with $\mathbf{s}$. A better alternative to this simple rounding is the Gaussian randomized rounding; see [2], [8], [9] for the details.

It has been shown that PI-SDR is a bidual of ML, a desirable property from a Lagrangian duality viewpoint. Specifically, (6) achieves the optimal Lagrangian dual lower bound of the polynomial ML formulation in (5).

*B. Bound constrained SDR*

BC-SDR is possibly the simplest among the various 16-QAM SDR methods. It relaxes the ML problem in (4) to an SDP

$$\begin{aligned} \min \quad & \text{tr}(\mathbf{H}^T\mathbf{H}\mathbf{S}) - 2\mathbf{s}^T\mathbf{H}^T\mathbf{y} + \|\mathbf{y}\|^2 \\ \text{s.t.} \quad & \mathbf{S} \succeq \mathbf{s}\mathbf{s}^T, \quad 1 \leq S_{ii} \leq 9, \quad i = 1, \ldots, N, \end{aligned} \quad (7)$$

where the original constraint $\mathbf{S} = \mathbf{s}\mathbf{s}^T$ is replaced by the PSD constraint $\mathbf{S} \succeq \mathbf{s}\mathbf{s}^T$ (as in PI-SDR), and the discrete set $\{1, 9\}$ is relaxed to an interval $[1, 9]$.

The BC-SDR problem in (7) exhibits particularly simple SDP problem structure. This has enabled us to develop a specialized interior-point algorithm for (7) that runs many times faster than some general-purpose interior-point software [21].

While PI-SDR is bidual of ML, it is not known if BC-SDR has such a desirable property.

*C. Virtually Antipodal SDR*

VA-SDR was proposed by Mao *et al.* [10][2] who also suggested a MIMO detector that combines VA-SDR and multistage decision-aided cancellation. Here we are interested only in VA-SDR. The idea stems from the fact that

$$s \in \{\pm 1, \pm 3\} \iff s = b_1 + 2b_2, \quad b_1, b_2 \in \{\pm 1\}.$$

Hence, the 16-QAM ML problem can be re-expressed in a virtually antipodal form

$$\min_{\mathbf{b}_1, \mathbf{b}_2 \in \{\pm 1\}^N} \|\mathbf{y} - \mathbf{H}(\mathbf{b}_1 + 2\mathbf{b}_2)\|^2 = \min_{\mathbf{b} \in \{\pm 1\}^{2N}} \|\mathbf{y} - \mathbf{H}\mathbf{W}\mathbf{b}\|^2, \quad (8)$$

where we denote

$$\mathbf{W} = [\ \mathbf{I}\ 2\mathbf{I}\ ], \qquad \mathbf{b} = [\ \mathbf{b}_1^T\ \mathbf{b}_2^T\ ]^T.$$

By applying the same SDR as in BPSK/QPSK constellations, VA-SDR is obtained:

$$\begin{aligned} \min \quad & \text{tr}(\mathbf{W}^T\mathbf{H}^T\mathbf{H}\mathbf{W}\mathbf{B}) - 2\mathbf{b}^T\mathbf{W}^T\mathbf{H}^T\mathbf{y} + \|\mathbf{y}\|^2 \\ \text{s.t.} \quad & \mathbf{B} \succeq \mathbf{b}\mathbf{b}^T, \quad B_{ii} = 1, \quad i = 1, \ldots, 2N. \end{aligned} \quad (9)$$

In terms of problem structure, VA-SDR is exactly the same as the SDR for BPSK/QPSK. Hence, VA-SDR can be implemented by directly applying interior-point algorithms designed for BPSK/QPSK SDR [20], [26].

---

[2]In fact, an earlier work by Steingrimsson *et al.* [7] was close to finding VA-SDR. In that paper, a symbol is considered as a linear transformation of some bit symbols, which is exactly how VA-SDR works. But we should emphasize that it was Mao *et al.* [10] who first described the use of VA-SDR for higher-order QAM and put the method to the test.

It can be shown that VA-SDR is also a bidual of ML. Specifically, (9) achieves the optimal Lagrangian dual lower bound of the virtually antipodal ML formulation in (8); see the literature such as [28, Section 5.15 and Exercise 5.39] for the proof. We however should point out that the biduality of VA-SDR and PI-SDR does not translate into equivalence of the two SDRs. The reason is that the two SDRs are obtained from two *different* ML formulations, which may be viewed as different optimization problems that have the same optimal value. The two ML formulations could yield different dual optimal values (and consequently different SDR optimal values), unless the two formulations have strong duality which is unlikely in this problem.

### D. Comparisons by Simulations

Let us use simulations to compare the above described SDR methods, before proceeding to analyzing them. The simulation setting follows that of a standard MIMO system, where the channel matrix $\tilde{\mathbf{H}}$ is i.i.d. complex circular Gaussian distributed with zero mean and unit variance. The MIMO system size is $(\tilde{M}, \tilde{N}) = (8, 8)$. For PI-SDR and BC-SDR, we employ the *simple rounding* procedure; i.e., if $\mathbf{s}^\star$ is the PI-SDR/BC-SDR solution associated with $\mathbf{s}$, then

$$\hat{\mathbf{s}} = \mathrm{dec}(\mathbf{s}^\star)$$

is the detected symbol vector where $\mathrm{dec}(\cdot)$ is the elementwise decision function for the discrete set $\{\pm 1, \pm 3\}$. For VA-SDR, there are two possible ways of doing simple rounding. Let $\mathbf{b}^\star = [\,(\mathbf{b}_1^\star)^T \,\, (\mathbf{b}_2^\star)^T\,]^T$ ($\mathbf{b}_1^\star, \mathbf{b}_2^\star \in \mathbb{R}^N$) be the VA-SDR solution associated with $\mathbf{b}$. We can detect $\mathbf{s}$ either by

$$\hat{\mathbf{s}} = \mathrm{sgn}(\mathbf{b}_1^\star) + 2\mathrm{sgn}(\mathbf{b}_2^\star) \tag{10}$$

where $\mathrm{sgn}(\cdot)$ denotes the elementwise sign function, or by

$$\hat{\mathbf{s}} = \mathrm{dec}(\mathbf{b}_1^\star + 2\mathbf{b}_2^\star) = \mathrm{dec}(\mathbf{W}\mathbf{b}^\star). \tag{11}$$

We call (10) and (11) *simple rounding I* and *simple rounding II*, respectively.

The simulated symbol error performance of the three SDRs is given in Fig. 1. In the figure the SNR is defined as the received signal-to-noise ratio per QAM symbol; i.e., $\mathrm{E}\{\|\tilde{\mathbf{h}}_i \tilde{s}_i\|^2\}/\mathrm{E}\{\|\tilde{\boldsymbol{\nu}}\|^2\}$. One can see that for VA-SDR, simple rounding II gives better performance than simple rounding I. But, more importantly, the performance of PI-SDR, BC-SDR, and VA-SDR (with simple rounding II) is identical. From this observation it is reasonable to suspect that there are strong connections between the three SDRs.

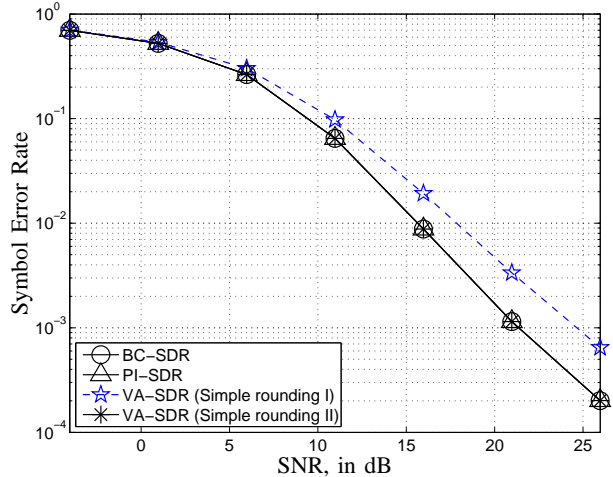

Fig. 1. Comparison of symbol error rates of PI-SDR, BC-SDR, and VA-SDR in an $8 \times 8$ 16-QAM system.

## IV. Equivalence of the Three 16-QAM SDR Detectors

In this section we prove the equivalence of PI-SDR, BC-SDR, and VA-SDR in the 16-QAM case. In the first subsection, the equivalence result and its implications will be described. Then, the analysis leading to the equivalence result will be shown in details in the second and third subsections.

### A. Main Result and Implications

The three SDRs can be represented by a unified expression

$$\min_{(\mathbf{S},\mathbf{s}) \in \mathcal{F}} f(\mathbf{S}, \mathbf{s}) \tag{12}$$

where

$$f(\mathbf{S}, \mathbf{s}) = \mathrm{tr}(\mathbf{H}^T \mathbf{H} \mathbf{S}) - 2\mathbf{s}^T \mathbf{H}^T \mathbf{y} + \|\mathbf{y}\|^2$$

is the objective function, and $\mathcal{F}$ is the feasible set, the definition of which depends on the SDR method employed. For BC-SDR, the feasible set is defined as

$$\mathcal{F}_{\mathsf{BC-SDR}} = \{\ (\mathbf{S}, \mathbf{s}) \mid \mathbf{S} \succeq \mathbf{s}\mathbf{s}^T, \mathbf{1}_N \preceq d(\mathbf{S}) \preceq 9\mathbf{1}_N\ \} \tag{13}$$

(We adopt the standard notation that '$\preceq$' and '$\succeq$' mean elementwise inequalities, when applied on vectors). For PI-SDR, the feasible set is characterized as

$$\mathcal{F}_{\mathsf{PI-SDR}} = \{\ (\mathbf{S}, \mathbf{s}) \mid (\mathbf{U}, \mathbf{u}, \mathbf{S}, \mathbf{s}) \in \mathcal{W}_{\mathsf{PI-SDR}}\ \} \tag{14}$$

$$\mathcal{W}_{\mathsf{PI-SDR}} = \{\ (\mathbf{U}, \mathbf{u}, \mathbf{S}, \mathbf{s}) \mid \mathbf{U} \succeq \mathbf{u}\mathbf{u}^T, \mathbf{S} \succeq \mathbf{s}\mathbf{s}^T, d(\mathbf{S}) = \mathbf{u}, d(\mathbf{U}) - 10\mathbf{u} + 9\mathbf{1}_N = \mathbf{0}\ \}, \tag{15}$$

and for VA-SDR,

$$\mathcal{F}_{\text{VA-SDR}} = \{ \ (\mathbf{S}, \mathbf{s}) = (\mathbf{W}\mathbf{B}\mathbf{W}^T, \mathbf{W}\mathbf{b}) \mid (\mathbf{B}, \mathbf{b}) \in \mathcal{B}_{\text{VA-SDR}} \ \} \tag{16}$$

$$\mathcal{B}_{\text{VA-SDR}} = \{ \ (\mathbf{B}, \mathbf{b}) \mid \mathbf{B} \succeq \mathbf{b}\mathbf{b}^T, d(\mathbf{B}) = \mathbf{1}_{2N} \ \}. \tag{17}$$

Essentially, the equivalence of the three SDRs lies in the feasible set:

**Theorem 1** *The feasible sets of the three 16-QAM SDRs are identical; that is,*

$$\mathcal{F}_{\text{PI-SDR}} = \mathcal{F}_{\text{BC-SDR}} = \mathcal{F}_{\text{VA-SDR}}.$$

The proof will be described in the next two subsections. From Theorem 1 we make the important conclusion that

**Corollary 1** *For 16-QAM MIMO detection, the relaxation problems of PI-SDR, BC-SDR, and VA-SDR [given in (6), (7), and (9), respectively] are equivalent. In particular,*

1) *if $(\tilde{\mathbf{U}}, \tilde{\mathbf{u}}, \tilde{\mathbf{S}}, \tilde{\mathbf{s}})$ is an optimal solution of PI-SDR, then $(\tilde{\mathbf{S}}, \tilde{\mathbf{s}})$ is an optimal solution of BC-SDR;*
2) *if $(\check{\mathbf{B}}, \check{\mathbf{b}})$ is an optimal solution of VA-SDR, then $(\mathbf{W}\check{\mathbf{B}}\mathbf{W}^T, \mathbf{W}\check{\mathbf{b}})$ is an optimal solution of BC-SDR;*
3) *if $(\mathbf{S}^\star, \mathbf{s}^\star)$ is an optimal solution of BC-SDR, then there exists $(\mathbf{U}^\star, \mathbf{u}^\star)$ such that $(\mathbf{U}^\star, \mathbf{u}^\star, \mathbf{S}^\star, \mathbf{s}^\star)$ is an optimal solution of PI-SDR; and*
4) *if $(\mathbf{S}^\star, \mathbf{s}^\star)$ is an optimal solution of BC-SDR, then there exists $(\mathbf{B}^\star, \mathbf{b}^\star)$ such that $(\mathbf{W}\mathbf{B}^\star\mathbf{W}^T, \mathbf{W}\mathbf{b}^\star) = (\mathbf{S}^\star, \mathbf{s}^\star)$ and $(\mathbf{B}^\star, \mathbf{b}^\star)$ is an optimal solution of VA-SDR.*

Some further discussions are as follows.

1) The SDR equivalence established here does not depend on the objective function $f$. Hence, the equivalence applies also to other similar SDR applications where the objective function is different. One such application is blind ML detection of orthogonal space-time block codes [22], in which $f$ is quasi-convex and is rather different from what is considered in this paper.
2) From Corollary 1 we see that an optimal BC-SDR solution can be directly obtained from an optimal PI-SDR or VA-SDR solution. In fact, an optimal PI-SDR or VA-SDR solution can also be constructed from an optimal BC-SDR solution in a closed-form manner. This is because our proof of Theorem 1 is constructive, which shows how a BC-SDR feasible point can be converted to a PI-SDR or VA-SDR feasible point (and vice versa). Simply speaking, constructing a PI-SDR solution is rather straightforward, and constructing a VA-SDR solution is more involved requiring some matrix factorization and vector decomposition procedures; for the construction details readers are referred to the proof in the following subsections.

3) The three SDRs can be proven to be equivalent for larger QAM sizes. For the equivalence of VA-SDR and BC-SDR, the proof can be generalized using a similar principle. But, for the equivalence of PI-SDR and BC-SDR, the proof is much harder and tedious even for 64-QAM. This will be elaborated upon in the next section.

The proof of Theorem 1 consists of two parts: proving that $\mathcal{F}_{\text{BC-SDR}} = \mathcal{F}_{\text{PI-SDR}}$, and $\mathcal{F}_{\text{BC-SDR}} = \mathcal{F}_{\text{VA-SDR}}$.

### B. First Part of the Proof of Theorem 1: $\mathcal{F}_{\text{BC-SDR}} = \mathcal{F}_{\text{PI-SDR}}$

We first show that if $(\mathbf{U}, \mathbf{u}, \mathbf{S}, \mathbf{s}) \in \mathcal{W}_{\text{PI-SDR}}$, then $(\mathbf{S}, \mathbf{s})$ is feasible to $\mathcal{F}_{\text{BC-SDR}}$. Given $(\mathbf{U}, \mathbf{u}, \mathbf{S}, \mathbf{s}) \in \mathcal{W}_{\text{PI-SDR}}$, the PI-SDR feasibility condition $\mathbf{U} \succeq \mathbf{u}\mathbf{u}^T$ implies that $U_{ii} \geq u_i^2$ for all $i = 1, \ldots, N$. Hence,

$$0 = U_{ii} - 10u_i + 9 \geq u_i^2 - 10u_i + 9 = (u_i - 1)(u_i - 9)$$

for all $i$. The inequality above is the same as $(S_{ii} - 1)(S_{ii} - 9) \leq 0$, or $1 \leq S_{ii} \leq 9$. This shows that $(\mathbf{S}, \mathbf{s}) \in \mathcal{F}_{\text{BC-SDR}}$.

Next, we show that for any $(\mathbf{S}, \mathbf{s}) \in \mathcal{F}_{\text{BC-SDR}}$, we can explicitly construct a $(\mathbf{U}, \mathbf{u})$ such that $(\mathbf{U}, \mathbf{u}, \mathbf{S}, \mathbf{s}) \in \mathcal{W}_{\text{PI-SDR}}$. Consider the following construction from $(\mathbf{S}, \mathbf{s}) \in \mathcal{F}_{\text{BC-SDR}}$:

$$\mathbf{u} = d(\mathbf{S})$$
$$\mathbf{U} = \mathbf{u}\mathbf{u}^T + D(\mathbf{w}) \tag{18}$$

where $D : \mathbb{R}^N \to \mathbb{R}^{N \times N}$ is the operator that outputs a diagonal matrix with its main diagonals being the input, and $\mathbf{w}$ is given by

$$w_i = -(S_{ii} - 1)(S_{ii} - 9) = -(u_i - 1)(u_i - 9), \tag{19}$$

for $i = 1, \ldots, N$. Since $1 \leq S_{ii} \leq 9$, we have $w_i \geq 0$. It follows that $\mathbf{U} - \mathbf{u}\mathbf{u}^T = D(\mathbf{w}) \succeq \mathbf{0}$. Moreover, from (18)-(19), one can see that

$$U_{ii} - 10u_i + 9 = w_i + u_i^2 - 10u_i + 9 = 0$$

for all $i$. This proves that $(\mathbf{U}, \mathbf{u}, \mathbf{S}, \mathbf{s})$ is feasible to $\mathcal{W}_{\text{PI-SDR}}$.

The proof above indicates that whatever a point is feasible to $\mathcal{F}_{\text{PI-SDR}}$ it is also feasible to $\mathcal{F}_{\text{BC-SDR}}$, and vice versa. We therefore conclude that $\mathcal{F}_{\text{BC-SDR}} = \mathcal{F}_{\text{PI-SDR}}$.

*C. Second Part of the Proof of Theorem 1:* $\mathcal{F}_{\text{VA-SDR}} = \mathcal{F}_{\text{BC-SDR}}$

Let $\mathbf{X} \in \mathbb{S}^{N+1}$ and $\mathbf{Y} \in \mathbb{S}^{2N+1}$ be two PSD matrices taking the form

$$\mathbf{X} = \begin{bmatrix} \mathbf{S} & \mathbf{s} \\ \mathbf{s}^T & 1 \end{bmatrix} \succeq \mathbf{0}, \qquad \mathbf{Y} = \begin{bmatrix} \mathbf{B} & \mathbf{b} \\ \mathbf{b}^T & 1 \end{bmatrix} \succeq \mathbf{0}$$

where $(\mathbf{S}, \mathbf{s}) \in \mathbb{S}^N \times \mathbb{R}^N$, $(\mathbf{B}, \mathbf{b}) \in \mathbb{S}^{2N} \times \mathbb{R}^{2N}$. By Schur complement, the two matrices satisfy $\mathbf{S} \succeq \mathbf{s}\mathbf{s}^T$ and $\mathbf{B} \succeq \mathbf{b}\mathbf{b}^T$. We assume $(\mathbf{S}, \mathbf{s}) = (\mathbf{W}\mathbf{B}\mathbf{W}^T, \mathbf{W}\mathbf{b})$, and this condition can be expressed in a matrix form

$$\mathbf{X} = \mathbf{T}^T \mathbf{Y} \mathbf{T} \tag{20}$$

where

$$\mathbf{T} = \begin{bmatrix} \mathbf{W}^T & \mathbf{0} \\ \mathbf{0} & 1 \end{bmatrix}.$$

Since $\mathbf{Y} \succeq \mathbf{0}$, $\mathbf{Y}$ can always be represented in a square-root factorization form

$$\mathbf{Y} = \mathbf{R}^T \mathbf{R}$$

for some square root factor $\mathbf{R} = [\, \mathbf{r}_1, \ldots, \mathbf{r}_{2N+1} \,] \in \mathbb{R}^{(2N+1) \times (2N+1)}$, with $\|\mathbf{r}_{2N+1}\| = 1$ (owing to $\|\mathbf{r}_i\|^2 = Y_{ii}$ and $Y_{2N+1, 2N+1} = 1$). Similarly, $\mathbf{X}$ can be characterized as

$$\mathbf{X} = \mathbf{Z}^T \mathbf{Z}$$

for some square root factor $\mathbf{Z} = [\, \mathbf{z}_1, \ldots, \mathbf{z}_{N+1} \,] \in \mathbb{R}^{(2N+1) \times (N+1)}$, $\|\mathbf{z}_{N+1}\| = 1$. We see that (20) holds if

$$\mathbf{Z} = \mathbf{R}\mathbf{T}. \tag{21}$$

Let us partition

$$\mathbf{R} = \begin{bmatrix} \mathbf{U} & \mathbf{V} & \mathbf{r}_{2N+1} \end{bmatrix} \tag{22}$$

where $\mathbf{U}, \mathbf{V} \in \mathbb{R}^{(2N+1) \times N}$. Substituting (22) into (21), we show that $\mathbf{Z} = [\, \mathbf{U} + 2\mathbf{V} \mid \mathbf{r}_{2N+1} \,]$, or equivalently

$$\mathbf{z}_i = \mathbf{u}_i + 2\mathbf{v}_i, \quad i = 1, \ldots, N, \tag{23}$$

$$\mathbf{z}_{N+1} = \mathbf{r}_{2N+1}, \tag{24}$$

where $\mathbf{u}_i$ is the $i$th column of $\mathbf{U}$, and $\mathbf{v}_i$ is defined in a similar way.

Now, suppose $(\mathbf{B}, \mathbf{b}) \in \mathcal{B}_{\mathsf{VA-SDR}}$. Since $d_i(\mathbf{B}) = 1$ for all $i$ (where $d_i(\cdot)$ means that $d_i(\mathbf{A}) = A_{ii}$), we have $\|\mathbf{u}_i\| = Y_{ii} = d_i(\mathbf{B}) = 1$ and $\|\mathbf{v}_i\| = Y_{i+N,i+N} = d_{i+N}(\mathbf{B}) = 1$ for $i = 1, \ldots, N$. With (23)-(24) satisfied, it holds true that

$$\|\mathbf{z}_i\| \leq \|\mathbf{u}_i\| + 2\|\mathbf{v}_i\| = 3,$$

$$\|\mathbf{z}_i\| \geq 2\|\mathbf{v}_i\| - \|\mathbf{u}_i\| = 1,$$

for $i = 1, \ldots, N$. This translates into an $\mathbf{S}$ that satisfies $d_i(\mathbf{S}) = X_{ii} = \|\mathbf{z}_i\|^2 \in [1, 9]$. And this further implies that $(\mathbf{S}, \mathbf{s}) \in \mathcal{F}_{\mathsf{BC-SDR}}$. On the other hand, suppose $(\mathbf{S}, \mathbf{s}) \in \mathcal{F}_{\mathsf{BC-SDR}}$. There is no problem for (24) to be satisfied, and we find $(\mathbf{u}_i, \mathbf{v}_i)$ satisfying (23) by resorting to the following lemma:

**Lemma 1** *Let $\mathbf{z} \in \mathbb{R}^n$, $n \geq 2$ be a given vector satisfying*

$$\beta - \alpha \leq \|\mathbf{z}\| \leq \beta + \alpha$$

*for some $\alpha, \beta > 0$. Then there exist two unit 2-norm vectors $\mathbf{u}$ and $\mathbf{v}$ such that*

$$\mathbf{z} = \alpha \mathbf{u} + \beta \mathbf{v}.$$

The proof of Lemma 1 is given in Appendix A. Essentially, the proof shows how to construct $(\mathbf{u}, \mathbf{v})$ from $\mathbf{z}$ in a closed-form manner. Applying Lemma 1 to (23) (with $\alpha = 1$ and $\beta = 2$), for each $i$ we obtain $(\mathbf{u}_i, \mathbf{v}_i)$ that satisfies (23) for any $\|\mathbf{z}_i\| \in [1, 3]$ (or $d_i(\mathbf{S}) \in [1, 9]$) and then achieves $\|\mathbf{u}_i\| = \|\mathbf{v}_i\| = 1$ at the same time. This means that the resultant $\mathbf{R}$ [cf., Eq. (22)] has unit 2-norm columns, and as a consequence $d_i(\mathbf{B}) = Y_{ii} = \|\mathbf{z}_i\|^2 = 1$. Hence, we have $(\mathbf{B}, \mathbf{b}) \in \mathcal{B}_{\mathsf{VA-SDR}}$.

We have shown by construction that $\mathcal{F}_{\mathsf{VA-SDR}} = \mathcal{F}_{\mathsf{BC-SDR}}$.

## V. GENERALIZATIONS TO LARGER QAM SIZES

Our endeavor now turns to more challenging cases in which the QAM size is larger than 16. In what follows, we will prove that i) for any $4^q$-QAM constellation (where $q \geq 1$), VA-SDR is equivalent to BC-SDR; and that ii) for the 64-QAM constellation, PI-SDR is equivalent to BC-SDR. Details regarding i) and ii) will be described in the first and second subsections, respectively.

### A. Equivalence of VA-SDR and BC-SDR for $4^q$-QAM

For $4^q$-QAM, the ML problem to be addressed is

$$\begin{aligned} \min \quad & \|\mathbf{y} - \mathbf{H}\mathbf{s}\|^2 \\ \text{s.t.} \quad & s_i \in \{\pm 1, \pm 3, \pm 5, \ldots, \pm(2^q - 1)\}, \quad i = 1, \ldots, N. \end{aligned}$$

Its virtually antipodal formulation takes the form

$$\min_{\mathbf{b} \in \{\pm 1\}^{qN}} \|\mathbf{y} - \mathbf{HWb}\|^2$$

where

$$\mathbf{W} = [\ \mathbf{I}\ 2\mathbf{I}\ 4\mathbf{I}\ 8\mathbf{I} \ldots\ 2^{q-1}\mathbf{I}\ ] \in \mathbb{R}^{N \times qN}$$

and

$$\mathbf{b} = [\ \mathbf{b}_1^T\ \mathbf{b}_2^T \ldots \mathbf{b}_q^T\ ]^T \in \mathbb{R}^{qN}$$

with $\mathbf{b}_i \in \mathbb{R}^N$ for all $i$. Again, both the VA-SDR and BC-SDR problems in this case can be represented by the expression

$$\min_{(\mathbf{S}, \mathbf{s}) \in \mathcal{F}} f(\mathbf{S}, \mathbf{s})$$

where the feasible set $\mathcal{F}$ for BC-SDR is defined as

$$\mathcal{F}_{\text{BC-SDR}} = \{\ (\mathbf{S}, \mathbf{s})\ |\ \mathbf{S} \succeq \mathbf{ss}^T, \mathbf{1}_N \preceq d(\mathbf{S}) \preceq (2^q - 1)^2 \mathbf{1}_N\ \} \quad (25)$$

and the feasible set for VA-SDR is

$$\mathcal{F}_{\text{VA-SDR}} = \{\ (\mathbf{S}, \mathbf{s}) = (\mathbf{WBW}^T, \mathbf{Wb})\ |\ (\mathbf{B}, \mathbf{b}) \in \mathcal{B}_{\text{VA-SDR}}\ \} \quad (26)$$

$$\mathcal{B}_{\text{VA-SDR}} = \{\ (\mathbf{B}, \mathbf{b})\ |\ \mathbf{B} \succeq \mathbf{bb}^T, d(\mathbf{B}) = \mathbf{1}_{qN}\ \} \quad (27)$$

It is shown that the equivalence of BC-SDR and VA-SDR is promised even for higher-order QAM.

**Theorem 2** *Consider a $4^q$-QAM constellation, where $q \geq 1$. It holds true that*

$$\mathcal{F}_{\text{VA-SDR}} = \mathcal{F}_{\text{BC-SDR}}.$$

The proof of Theorem 2 is given in Appendix B. It is a generalization of its 16-QAM counterpart in Section IV-C. Like the 16-QAM case, the proof reveals the possibility that an optimal BC-SDR solution can be used to construct an optimal VA-SDR solution in an analytical fashion, or vice versa.

*B. Equivalence of PI-SDR and BC-SDR for $64$-QAM*

The original work of PI-SDR [8] concentrates only on the 16-QAM constellation, but it is clear from that work that the idea can be extended to the 64-QAM constellation. To see this, we start with the following 64-QAM ML formulation

$$\begin{aligned}\min\quad & \|\mathbf{y} - \mathbf{Hs}\|^2 \\ \text{s.t.}\quad & s_i^2 \in \{r_1, r_2, r_3, r_4\}, \quad i = 1, \ldots, N\end{aligned}$$

where $\{r_1, r_2, r_3, r_4\} = \{1, 3^2, 5^2, 7^2\}$. The idea is to consider the polynomial characterization

$$u \in \{r_1, r_2, r_3, r_4\} \iff \prod_{i=1}^{4}(u - r_i) = \sum_{\ell=1}^{5} p_\ell u^{\ell-1} = 0$$

where $\{p_i\}$ is the set of polynomial coefficients associated with the roots $\{r_i\}$. Like the development in 16-QAM PI-SDR, we reformulate the ML problem as

$$\begin{aligned} \min_{\mathbf{U},\mathbf{u},\mathbf{S},\mathbf{s}} \quad & f(\mathbf{S},\mathbf{s}) \\ \text{s.t.} \quad & \mathbf{U} = \begin{bmatrix} \mathbf{U}_{11} & \mathbf{U}_{12} \\ \mathbf{U}_{12}^T & \mathbf{U}_{22} \end{bmatrix}, \mathbf{u} = \begin{bmatrix} \mathbf{u}_1 \\ \mathbf{u}_2 \end{bmatrix} \\ & \mathbf{S} = \mathbf{s}\mathbf{s}^T, \mathbf{U} = \mathbf{u}\mathbf{u}^T \\ & d(\mathbf{S}) = \mathbf{u}_1, d(\mathbf{U}_{11}) = \mathbf{u}_2 \\ & p_1 \mathbf{1}_N + p_2 \mathbf{u}_1 + p_3 d(\mathbf{U}_{11}) + p_4 d(\mathbf{U}_{12}) + p_5 d(\mathbf{U}_{22}) = \mathbf{0}. \end{aligned} \quad (28)$$

The formation in (28) is valid because its constraints essentially restrict $u_i = s_i^2$, $d_i(\mathbf{U}_{11}) = u_i^2$, $d_i(\mathbf{U}_{12}) = u_i^3$, and $d_i(\mathbf{U}_{22}) = u_i^4$. From (28), we obtain the 64-QAM PI-SDR:

$$\begin{aligned} \min_{\mathbf{U},\mathbf{u},\mathbf{S},\mathbf{s}} \quad & f(\mathbf{S},\mathbf{s}) \\ \text{s.t.} \quad & \mathbf{U} = \begin{bmatrix} \mathbf{U}_{11} & \mathbf{U}_{12} \\ \mathbf{U}_{12}^T & \mathbf{U}_{22} \end{bmatrix}, \mathbf{u} = \begin{bmatrix} \mathbf{u}_1 \\ \mathbf{u}_2 \end{bmatrix} \\ & \mathbf{S} \succeq \mathbf{s}\mathbf{s}^T, \mathbf{U} \succeq \mathbf{u}\mathbf{u}^T \\ & d(\mathbf{S}) = \mathbf{u}_1, d(\mathbf{U}_{11}) = \mathbf{u}_2 \\ & p_1 \mathbf{1}_N + p_2 \mathbf{u}_1 + p_3 d(\mathbf{U}_{11}) + p_4 d(\mathbf{U}_{12}) + p_5 d(\mathbf{U}_{22}) = \mathbf{0}. \end{aligned} \quad (29)$$

For BC-SDR, the relaxation is given by

$$\begin{aligned} \min \quad & f(\mathbf{S},\mathbf{s}) \\ \text{s.t.} \quad & \mathbf{S} \succeq \mathbf{s}\mathbf{s}^T, \quad r_1 \mathbf{1}_N \preceq d(\mathbf{S}) \preceq r_4 \mathbf{1}_N. \end{aligned} \quad (30)$$

The main result here is presented as follows:

**Theorem 3** *Consider a general situation where the roots $\{r_i\}$ are allowed to be arbitrary (not necessarily the roots in 64-QAM), and assume $0 < r_1 < \ldots < r_4 < \infty$. The PI-SDR problem in (29) and the BC-SDR problem in (30) are equivalent in yielding the same feasible set corresponding to $(\mathbf{S}, \mathbf{s})$ (and thus the same optimal solutions), under the following sufficient and necessary condition*

$$\sqrt{r_4 - r_1} \leq \min\{\sqrt{r_3 - r_1} + \sqrt{r_2 - r_1}, \sqrt{r_4 - r_2} + \sqrt{r_4 - r_3}\}. \quad (31)$$

It can be verified that the 64-QAM roots ($\{r_1, r_2, r_3, r_4\} = \{1, 3^2, 5^2, 7^2\}$) satisfy (31). We therefore conclude that **Corollary 2** *For the 64-QAM constellation, the PI-SDR problem in* (29) *and the BC-SDR problem in* (30) *are equivalent in yielding the same feasible set corresponding to* $(\mathbf{S}, \mathbf{s})$.

*Proof of Theorem 3:* The proof is far from trivial compared to its 16-QAM counterpart. Consider the following lemma shown in Appendix C:

**Lemma 2** *The PI-SDR problem in* (29) *is equivalent to the following alternate PI-SDR problem*

$$\begin{aligned}
\min_{\mathbf{V}_1,\ldots,\mathbf{V}_N,\mathbf{S},\mathbf{s}} \quad & f(\mathbf{S},\mathbf{s}) \\
\text{s.t.} \quad & \mathbf{S} \succeq \mathbf{ss}^T, d(\mathbf{S}) = [\, v_{11},\ldots,v_{N,1}\, ]^T \\
& \mathbf{V}_i = \begin{bmatrix} 1 & v_{i1} & v_{i2} \\ v_{i1} & v_{i2} & v_{i3} \\ v_{i2} & v_{i3} & v_{i4} \end{bmatrix} \succeq \mathbf{0}, \quad i=1,\ldots,N \\
& p_1 + \sum_{\ell=1}^{4} p_{\ell+1} v_{i,\ell} = 0, \quad i=1,\ldots,N
\end{aligned} \quad (32)$$

*in the sense that the feasible sets corresponding to* $(\mathbf{S}, \mathbf{s})$ *are identical for the two problems.*

The proof of Lemma 2 follows the same approach as the equivalence proof for the 16-QAM PI-SDR and BC-SDR (in Section IV-B). However, by Lemma 2 alone, we are unable to see the equivalence of the 64-QAM PI-SDR and BC-SDR immediately.

To gain further insights, let us re-express the alternate PI-SDR formulation in (32) as

$$\begin{aligned}
\min \quad & f(\mathbf{S},\mathbf{s}) \\
\text{s.t.} \quad & \mathbf{S} \succeq \mathbf{ss}^T, S_{ii} \in \mathcal{D}, \quad i=1,\ldots,N
\end{aligned}$$

where we define

$$\mathcal{D} = \{\, S \in \mathbb{R} \mid S = [\mathbf{V}]_{12}, \mathbf{V} \in \mathcal{V}\, \} \quad (33)$$

$$\mathcal{V} = \left\{ \mathbf{V} \in \mathbb{S}^3 \,\middle|\, \mathbf{V} \succeq \mathbf{0}, \mathbf{V} = \operatorname{Hank}((1,\mathbf{v})), p_1 + \sum_{\ell=1}^{4} p_{\ell+1} v_\ell = 0, \mathbf{v} \in \mathbb{R}^4 \right\} \quad (34)$$

with the operator $\operatorname{Hank}: \mathbb{R}^{2n-1} \to \mathbb{R}^{n \times n}$ standing for

$$\operatorname{Hank}(a_1,\ldots,a_{2n-1}) = \begin{bmatrix} a_1 & a_2 & \cdots & a_n \\ a_2 & \ddots & & \vdots \\ \vdots & & \ddots & a_{2n-2} \\ a_n & \cdots & a_{2n-2} & a_{2n-1} \end{bmatrix}.$$

Our interest now turns to analyzing the set $\mathcal{D}$, which has to be done by analyzing $\mathcal{V}$. Consider the following lemma proven in Appendix D:

**Lemma 3** *The set $\mathcal{V}$ in* (34) *is equivalent to*

$$\mathcal{V} = \left\{ \mathbf{V} \in \mathbb{S}^3 \ \bigg| \ \mathbf{V} = \sum_{\ell=1}^{4} \theta_\ell \mathbf{a}_\ell \mathbf{a}_\ell^T, \mathbf{V} \succeq \mathbf{0}, \sum_{\ell=1}^{4} \theta_\ell = 1 \right\} \qquad (35)$$

*where* $\mathbf{a}_\ell = [\ 1\ r_\ell\ r_\ell^2\ ]^T$.

Lemma 3 provides an interesting implication. To describe it, let

$$\mathrm{conv}\{\mathbf{a}_1 \mathbf{a}_1^T, \ldots, \mathbf{a}_4 \mathbf{a}_4^T\} = \left\{ \mathbf{V} \ \bigg| \ \mathbf{V} = \sum_{\ell=1}^{4} \theta_\ell \mathbf{a}_\ell \mathbf{a}_\ell^T, \boldsymbol{\theta} \succeq \mathbf{0}, \sum_{\ell=1}^{4} \theta_\ell = 1 \right\}$$

be the convex hull of $\{\mathbf{a}_1 \mathbf{a}_1^T, \ldots, \mathbf{a}_4 \mathbf{a}_4^T\}$. It can be verified from (35) that $\mathcal{V} \supseteq \mathrm{conv}\{\mathbf{a}_1 \mathbf{a}_1^T, \ldots, \mathbf{a}_4 \mathbf{a}_4^T\}$, though $\mathcal{V} \subseteq \mathrm{conv}\{\mathbf{a}_1 \mathbf{a}_1^T, \ldots, \mathbf{a}_4 \mathbf{a}_4^T\}$ is generally not true[3]. Consequently, we have

$$\mathcal{D} \supseteq \{\ S = [\mathbf{V}]_{12}\ |\ \mathbf{V} \in \mathrm{conv}\{\mathbf{a}_1 \mathbf{a}_1^T, \ldots, \mathbf{a}_4 \mathbf{a}_4^T\}\ \}$$

$$= \left\{\ S = \sum_{\ell=1}^{4} \theta_\ell [\mathbf{a}_\ell \mathbf{a}_\ell^T]_{12}\ \bigg|\ \boldsymbol{\theta} \succeq \mathbf{0}, \sum_{\ell=1}^{4} \theta_\ell = 1\ \right\}$$

$$= \left\{ S = \sum_{\ell=1}^{4} \theta_\ell r_\ell\ \bigg|\ \boldsymbol{\theta} \succeq \mathbf{0}, \sum_{\ell=1}^{4} \theta_\ell = 1\ \right\} = [r_1, r_4].$$

*This implies that the 64-QAM PI-SDR is no tighter than the 64-QAM BC-SDR.* But we also show in Appendix E that

**Lemma 4** *Let* $0 < r_1 < \ldots < r_4 < \infty$. *We have* $\mathcal{D} = [r_1, r_4]$ *if and only if*

$$\sqrt{r_4 - r_1} \leq \min\{\sqrt{r_3 - r_1} + \sqrt{r_2 - r_1}, \sqrt{r_4 - r_2} + \sqrt{r_4 - r_3}\}.$$

As a result, PI-SDR can be equivalent to BC-SDR under the condition in Lemma 4, thereby completing the proof of Theorem 3.

## VI. SOME FURTHER SIMULATION RESULTS

We provide two more sets of simulation results. In the first set, the purpose is to verify the SDR equivalence for the 64-QAM and 256-QAM cases. The simulation settings are the same as those of the 16-QAM simulation example in Section III-D, and the MIMO size is $(\tilde{M}, \tilde{N}) = (4, 4)$. Simple rounding is employed for the SDR methods. The results are plotted in Fig. 2. We see that the symbol error rates (SERs) of the PI-SDR, BC-SDR, and VA-SDR with simple rounding II are generally identical, which corroborates our theoretical results. It is also noticed that the performance of 64-QAM PI-SDR slightly deviates from that of 64-QAM BC-SDR and VA-SDR at SNR= 45dB. We found that this was due to some numerical problems encountered by the interior-point SDP solver (which is SeDuMi [27] here). In fact, the polynomial coefficients in 64-QAM PI-SDR have values ranging from $p_5 = 1$ to $p_1 = 1 \times 3^2 \times 5^2 \times 7^2 = 11,025$. Such a large dynamic range could be the cause of the numerical inaccuracy. Moreover, Fig. 2 illustrates that VA-SDR with simple rounding I is not working in 64-QAM and 256-QAM (see Section III-D for the definition of simple roundings I and II). This problem, which has also been noticed by Mao *et al.* [10], may partially be answered by the equivalence proof for BC-SDR and VA-SDR; cf., Section IV-C and Appendix B. In essence, the derivations there revealed that the VA-SDR solution with respect to $(\mathbf{B}, \mathbf{b})$ may be non-unique, even though its BC-SDR counterpart [in form of $(\mathbf{S}, \mathbf{s})$] is unique. In particular, a key component, namely Lemma 1 is not a unique decomposition.

---

[3] By numerical test, it was found that there exists a $\mathbf{V} \in \mathcal{V}$ such that some of the constituent $\theta_\ell$ can be negative.

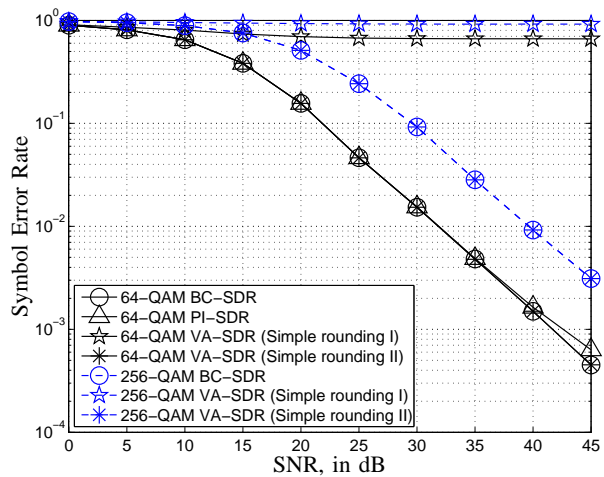

Fig. 2. Comparison of symbol error rates of PI-SDR, BC-SDR, and VA-SDR in an $4 \times 4$ system with either 64-QAM or 256-QAM.

In the second set of simulation results, a comparison is made between SDR and some other MIMO detectors for the 64-QAM case. Those detectors include the zero-forcing (ZF) detector, the optimal sphere decoder, the lattice reduction aided ZF (LRA-ZF) detector [29], [30]. It is worthwhile to mention that the LRA-ZF detector has been shown to achieve the full receive diversity [31]. We tested the BC-SDR method only, as the other two SDR methods will provide similar results. A specialized interior-point SDP solver [21] is used to handle the BC-SDR optimization. For solution rounding, we employ a better procedure, namely the Gaussian randomized rounding. The procedure has been described in [8], and the number of randomizations is set to $100$. The results are plotted in Fig. 3 (a similar set of results for 16-QAM is also available in [21]). Fig. 3(a) shows the case of $\tilde{M} = \tilde{N} = 8$, where we see that LRA-ZF detector gives better performance than BC-SDR except for some rather low SNR points. We increase the problem size to $\tilde{M} = \tilde{N} = 16$ in Fig. 3(b). For this problem size it is computationally too hard to run the optimal sphere decoder. The figure illustrates that for SNRs less than 27dB, BC-SDR outperforms LRA-ZF. Let us further increase the problem size to $\tilde{M} = \tilde{N} = 40$. As illustrated in Fig. 3(c), now BC-SDR exhibits further improved performance compared to LRA-ZF.

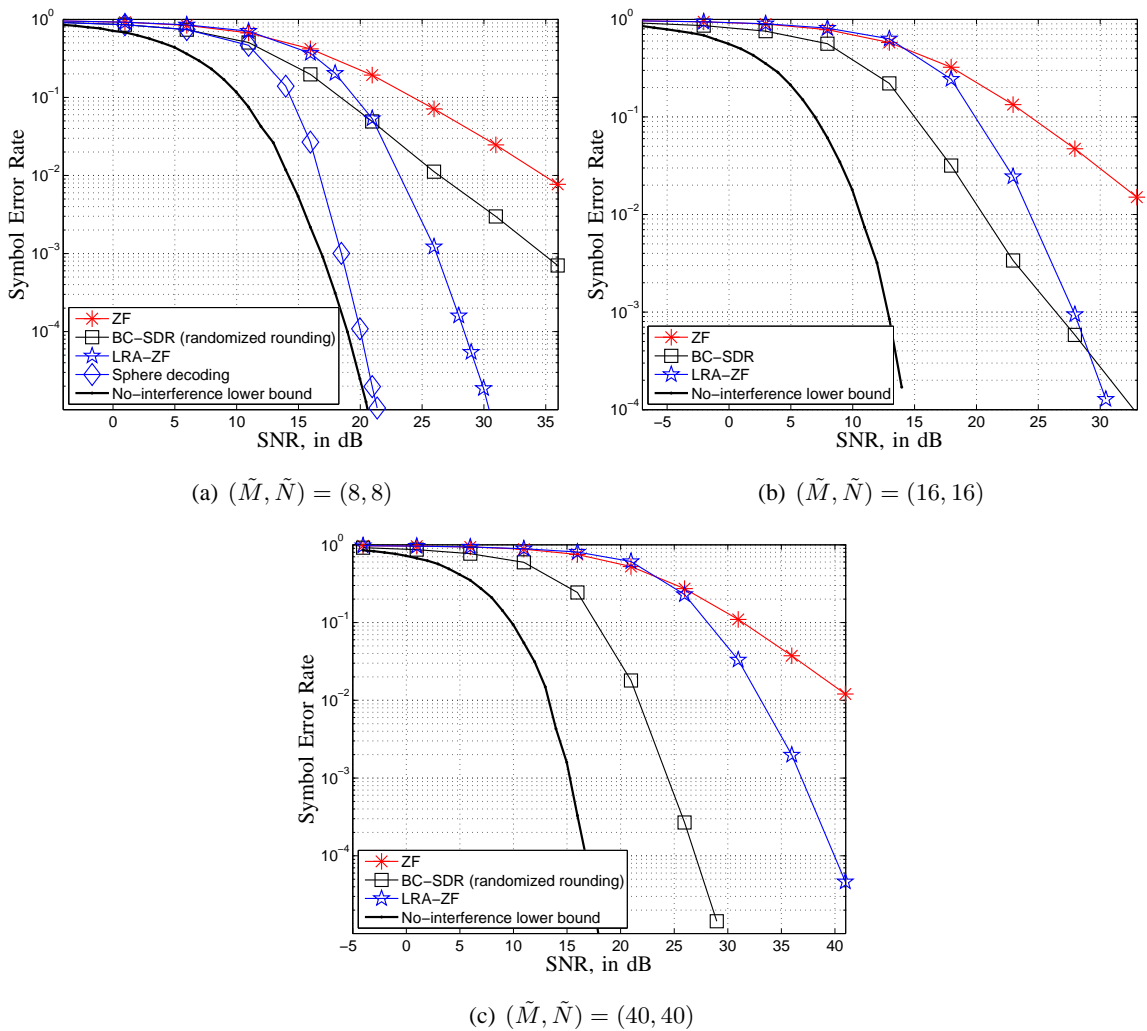

Fig. 3. Comparison of the various detectors in 64-QAM systems.

The comparisons above somehow suggests that SDR has significant performance advantages for large problem sizes and/or for low to moderate SNRs.

## VII. CONCLUSION AND DISCUSSION

This paper analyzes the relationships of three SDR-based MIMO detection methods for high-order QAM, namely PI-SDR, BC-SDR, and VA-SDR. We have proven that the three SDRs are actually equivalent, despite their different appearances. The essence of the equivalence is that an optimal solution of one SDR can always be constructed from that of another SDR. The equivalence result is not restricted to MIMO detection. It can be applied to other problems such as blind ML detection of orthogonal space-time block codes [22].

Since the three SDRs are now known to be equivalent, the comparison should turn to their computational costs. One can check numerically that BC-SDR is the cheapest computationally, followed by VA-SDR and then PI-SDR. Hence, it sounds that BC-SDR should be the method of choice among the three methods. While this is our present conclusion, our opinion is that each of the three SDRs is interesting in its own right where a different way of utilizing QAM structures has been demonstrated. Moreover, the exposition of the PI-SDR and VA-SDR ideas might help inspire future works for devising stronger SDR methods.

In the introduction we have mentioned the work by Yang *et al.* [11], who proposed an improved 16-QAM BC-SDR method by tightening the feasible set with more constraints. That work showed that the tightened BC-SDR method can perform better than the 16-QAM PI-SDR. Using the equivalence result here, one can actually infer that the tightened BC-SDR can perform better than the 16-QAM VA-SDR as well. But, presently it is not known if there is any connection of the above-described SDR methods to the SDR work by Mobasher *et al.* [12]. Studying this would be an interesting future direction.

## APPENDIX

### A. Proof of Lemma 1

The proof is constructive. Let

$$\mathbf{u} = \theta \frac{\mathbf{z}}{\|\mathbf{z}\|} + \sqrt{1-\theta^2}\mathbf{z}_\perp \qquad (36)$$

where $\mathbf{z}_\perp$ is a unit 2-norm vector orthogonal to $\mathbf{z}$, and $|\theta| \leq 1$. It can be verified that the $\mathbf{u}$ in (36) satisfies $\|\mathbf{u}\| = 1$. Moreover, let

$$\mathbf{v} = \frac{1}{\beta}(\mathbf{z} - \alpha\mathbf{u}) \qquad (37)$$

which is purposely constructed to satisfy $\mathbf{z} = \alpha\mathbf{u} + \beta\mathbf{v}$. One can show from (37) that the unit norm condition $\|\mathbf{v}\| = 1$ is achieved when we choose

$$\theta = \frac{\|\mathbf{z}\|^2 - (\beta^2 - \alpha^2)}{2\alpha\|\mathbf{z}\|}. \qquad (38)$$

Now the remaining problem is the condition under which $|\theta| \leq 1$. It can be verified from (38) that if $\beta - \alpha \leq \|\mathbf{z}\| \leq \beta + \alpha$ then $|\theta| \leq 1$ is guaranteed.

## B. Proof of Theorem 2

The idea is similar to the proof in the 16-QAM case, described in Section IV-C. We consider two PSD matrices

$$\mathbf{X} = \begin{bmatrix} \mathbf{S} & \mathbf{s} \\ \mathbf{s}^T & 1 \end{bmatrix} \succeq \mathbf{0}, \qquad \mathbf{Y} = \begin{bmatrix} \mathbf{B} & \mathbf{b} \\ \mathbf{b}^T & 1 \end{bmatrix} \succeq \mathbf{0}$$

that satisfy $(\mathbf{S}, \mathbf{b}) = (\mathbf{W}\mathbf{B}\mathbf{W}^T, \mathbf{W}\mathbf{b})$. That condition is shown to be achievable if

$$\mathbf{Z} = \mathbf{R} \begin{bmatrix} \mathbf{W}^T & \mathbf{0} \\ \mathbf{0} & 1 \end{bmatrix} \qquad (39)$$

where $\mathbf{Z} \in \mathbb{R}^{(qN+1) \times (N+1)}$ and $\mathbf{R} \in \mathbb{R}^{(qN+1) \times (qN+1)}$ are square root factors of $\mathbf{X}$ and $\mathbf{Y}$, respectively (or $\mathbf{Z}^T\mathbf{Z} = \mathbf{X}$, $\mathbf{R}^T\mathbf{R} = \mathbf{Y}$). The objective is to show that if $(\mathbf{S}, \mathbf{s}) \in \mathcal{F}_{\mathsf{BC-SDR}}$, then we can construct a $(\mathbf{B}, \mathbf{b}) \in \mathcal{B}_{\mathsf{VA-SDR}}$ satisfying (39); and vice versa.

For general $4^q$-QAM where $\mathbf{W}$ is expanded to $[\mathbf{I}\ 2\mathbf{I} \ldots 2^{q-1}\mathbf{I}]$, Eq. (39) can be rewritten as

$$\mathbf{z}_i = \sum_{j=0}^{q-1} 2^j \mathbf{r}_{i+jN}, \quad i = 1, \ldots, N, \qquad (40)$$

$$\mathbf{z}_{N+1} = \mathbf{r}_{qN+1}. \qquad (41)$$

To achieve (41) is easy, and the nontrivial part lies in (40). First, suppose $(\mathbf{B}, \mathbf{b}) \in \mathcal{B}_{\mathsf{VA-SDR}}$. Then the resultant $\mathbf{R}$ satisfies $\|\mathbf{r}_i\| = 1$. The vectors $\mathbf{z}_i$ satisfying (40) would then have bounds

$$\|\mathbf{z}_i\| \leq \sum_{j=0}^{q-1} 2^j \|\mathbf{r}_{i+jN}\| = \sum_{j=0}^{q-1} 2^j = 2^q - 1,$$

$$\|\mathbf{z}_i\| \geq 2^{q-1} \|\mathbf{r}_{i+(q-1)N}\| - \left\|\sum_{j=0}^{q-2} 2^j \mathbf{r}_{i+jN}\right\|$$

$$\geq 2^{q-1} - \sum_{j=0}^{q-2} 2^j = 2^{q-1} - (2^{q-1} - 1) = 1.$$

This means that the corresponding $\mathbf{S}$ has $d_i(\mathbf{S}) = \|\mathbf{z}_i\|^2 \in [1, (2^q - 1)^2]$. Hence, $(\mathbf{S}, \mathbf{s}) \in \mathcal{F}_{\mathsf{BC-SDR}}$. Second, suppose $(\mathbf{S}, \mathbf{s}) \in \mathcal{B}_{\mathsf{BC-SDR}}$. Let us choose, for each $i = 1, \ldots, N$,

$$\mathbf{r}_i = \mathbf{r}_{i+N} = \cdots = \mathbf{r}_{i+(q-2)N} \triangleq \mathbf{u}_i,$$

$$\mathbf{r}_{i+(q-1)N} \triangleq \mathbf{v}_i,$$

for some $\mathbf{u}_i, \mathbf{v}_i \in \mathbb{R}^{qN+1}$. The condition in (40) becomes

$$\mathbf{z}_i = \sum_{j=0}^{q-2} 2^j \mathbf{u}_i + 2^{q-1}\mathbf{v}_i = (2^{q-1}-1)\mathbf{u}_i + 2^{q-1}\mathbf{v}_i. \quad (42)$$

Using Lemma 1, we can construct a $(\mathbf{u}_i, \mathbf{v}_i)$ that satisfies (42) for any $\|\mathbf{z}_i\| = \sqrt{d_i(\mathbf{S})} \in [1, 2^q - 1]$ while achieving $\|\mathbf{u}_i\| = \|\mathbf{v}_i\| = 1$. The resultant $\mathbf{B}$ will therefore satisfy $d_i(\mathbf{B}) = Y_{ii} = 1$ for all $i$, meaning that $(\mathbf{B}, \mathbf{b}) \in \mathcal{B}_{\text{VA-SDR}}$.

The proof of Theorem 2 is complete.

*C. Proof of Lemma 2*

Suppose that $(\mathbf{U}, \mathbf{u}, \mathbf{S}, \mathbf{s})$ is feasible to the original 64-QAM PI-SDR problem in (29). Set

$$\mathbf{V}_i = \begin{bmatrix} 1 & \mathbf{0} \\ 0 & \mathbf{e}_i^T \\ 0 & \mathbf{e}_{i+N}^T \end{bmatrix} \begin{bmatrix} 1 & \mathbf{u}^T \\ \mathbf{u} & \mathbf{U} \end{bmatrix} \begin{bmatrix} 1 & 0 & 0 \\ \mathbf{0} & \mathbf{e}_i & \mathbf{e}_{i+N} \end{bmatrix} \quad (43)$$

for $i = 1, \ldots, N$, where $\mathbf{e}_i \in \mathbb{R}^{2N}$ is a unit vector with the nonzero element at the $i$th element. It follows from $\mathbf{U} \succeq \mathbf{u}\mathbf{u}^T$ and (43) that $\mathbf{V}_i \succeq \mathbf{0}$ for all $i$. Moreover, (43) equals

$$\mathbf{V}_i = \begin{bmatrix} 1 & u_{1i} & u_{2i} \\ u_{1i} & [\mathbf{U}_{11}]_{ii} & [\mathbf{U}_{12}]_{ii} \\ u_{2i} & [\mathbf{U}_{12}]_{ii} & [\mathbf{U}_{22}]_{ii} \end{bmatrix}. \quad (44)$$

Since $u_{2i} = [\mathbf{U}_{11}]_{ii}$, every $\mathbf{V}_i$ in (44) satisfies the Hankel structure in the alternate PI-SDR problem in (32). It also follows from (44) that the equality constraints arising from the polynomials are satisfied. Hence, $(\mathbf{V}_1, \ldots, \mathbf{V}_N, \mathbf{S}, \mathbf{s})$ is feasible to the alternate PI-SDR problem in (32).

On the other hand, suppose that $(\mathbf{V}_1, \ldots, \mathbf{V}_N, \mathbf{S}, \mathbf{s})$ is feasible to the alternate PI-SDR. Set

$$\mathbf{u}_1 = [\, v_{11}, \ldots, v_{N,1} \,]^T, \quad (45a)$$

$$\mathbf{U}_{11} = \mathbf{C}_2 - D(\mathbf{u}_1 \odot \mathbf{u}_1) + \mathbf{u}_1\mathbf{u}_1^T, \quad (45b)$$

$$\mathbf{u}_2 = d(\mathbf{U}_{11}) = [\, v_{12}, \ldots, v_{N,2} \,]^T, \quad (45c)$$

$$\mathbf{U}_{12} = \mathbf{C}_3 - D(\mathbf{u}_1 \odot \mathbf{u}_2) + \mathbf{u}_1\mathbf{u}_2^T, \quad (45d)$$

$$\mathbf{U}_{22} = \mathbf{C}_4 - D(\mathbf{u}_2 \odot \mathbf{u}_2) + \mathbf{u}_2\mathbf{u}_2^T, \quad (45e)$$

where $\odot$ is the Hadamard product, and

$$\mathbf{C}_i = D(v_{1,i}, \ldots, v_{N,i}).$$

It can be shown that (45) satisfies the equality constraints of the polynomials. Let us examine if $\mathbf{U} \succeq \mathbf{u}\mathbf{u}^T$. We see that

$$\mathbf{U} - \mathbf{u}\mathbf{u}^T = \begin{bmatrix} \mathbf{C}_2 - D(\mathbf{u}_1 \odot \mathbf{u}_1) & \mathbf{C}_3 - D(\mathbf{u}_1 \odot \mathbf{u}_2) \\ \mathbf{C}_3 - D(\mathbf{u}_1 \odot \mathbf{u}_2) & \mathbf{C}_4 - D(\mathbf{u}_2 \odot \mathbf{u}_2) \end{bmatrix}.$$

It can be shown using basic matrix results that the specially structured matrix above is PSD if and only if

$$\begin{bmatrix} v_{i,2} - u_{1,i}^2 & v_{i,3} - u_{1,i}u_{2,i} \\ v_{i,3} - u_{1,i}u_{2,i} & v_{i,4} - u_{2,i}^2 \end{bmatrix} \tag{46}$$

are PSD for all $i = 1, \ldots, N$. By using $u_{1,i} = v_{i,1}$ and $u_{2,i} = v_{i,2}$ and Schur complement, it is shown that (46) are indeed PSD. We therefore conclude that $(\mathbf{V}_1, \ldots, \mathbf{V}_N, \mathbf{S}, \mathbf{s})$ is feasible to the alternate PI-SDR problem.

*D. Proof of Lemma 3*

Consider a problem setting as follows: Let $\{r_1, \ldots, r_L\}$ be a given set of roots, and assume that the roots are distinct. Consider the following two sets

$$\mathcal{V}_1 = \left\{ \mathbf{V} = \text{Hank}(\mathbf{v}) \,\Big|\, \mathbf{v} \in \mathbb{R}^{L+1}, v_1 = 1, \mathbf{p}^T \mathbf{v} = 0, \mathbf{V} \succeq \mathbf{0} \right\}$$

where $\mathbf{p} = [\, p_1, \ldots, p_{L+1} \,]^T$ contains the polynomial coefficients corresponding to $\{r_1, \ldots, r_L\}$, i.e., $\sum_{\ell=1}^{L+1} p_\ell r^{\ell-1} = 0$ for all $r \in \{r_1, \ldots, r_L\}$; and

$$\mathcal{V}_2 = \left\{ \mathbf{V} = \sum_{\ell=1}^{L} \theta_\ell \mathbf{a}_\ell \mathbf{a}_\ell^T \,\Big|\, \sum_{\ell=1}^{L} \theta_\ell = 1, \mathbf{V} \succeq \mathbf{0} \right\}$$

where $\mathbf{a}_\ell = [\, 1 \; r_\ell \; r_\ell^2 \ldots r_\ell^{L/2} \,] \in \mathbb{R}^{L/2+1}$, and $L$ is assumed to be even. Our objective is to prove that $\mathcal{V}_1 = \mathcal{V}_2$. Clearly, Lemma 3 is a special case of the above problem where $L = 4$.

By definition, every $\mathbf{V} \in \mathcal{V}_1$ can be parameterized by some $\mathbf{v} \in \mathbb{R}^{L+1}$ such that $v_1 = 1$ and $\mathbf{p}^T \mathbf{v} = 0$. Let

$$\mathbf{b}_\ell = [\, 1, r_\ell, r_\ell^2, \ldots, r_\ell^L \,] \in \mathbb{R}^{L+1}$$

for $\ell = 1, \ldots, L$. Since every $\mathbf{b}_\ell$ contains one of the true roots, it satisfies $\mathbf{p}^T \mathbf{b}_\ell = 0$. Hence, we have the following condition to satisfy

$$\mathbf{p}^T [\, \mathbf{b}_1 \ldots \mathbf{b}_L \; \mathbf{v} \,] = \mathbf{0}. \tag{47}$$

The submatrix $[\,\mathbf{b}_1 \ldots \mathbf{b}_L\,] \in \mathbb{R}^{(L+1)\times L}$ is linearly independent, being Vandermonde with distinct roots. Subsequently, (47) can be satisfied only when

$$\mathbf{v} = \sum_{\ell=1}^{L} \theta_\ell \mathbf{b}_\ell.$$

for some coefficients $\boldsymbol{\theta} \in \mathbb{R}^L$. Since $1 = v_1 = \sum_{\ell=1}^{L} \theta_\ell [\mathbf{b}_\ell]_1 = \sum_{\ell=1}^{L} \theta_\ell$, the coefficients satisfy $\sum_{\ell=1}^{L} \theta_\ell = 1$. Moreover, by noticing that

$$\mathrm{Hank}(\mathbf{b}_\ell) = \mathbf{a}_\ell \mathbf{a}_\ell^T, \tag{48}$$

we have

$$\mathbf{V} = \mathrm{Hank}(\mathbf{v}) = \sum_{\ell=1}^{L} \theta_\ell \mathbf{a}_\ell \mathbf{a}_\ell^T.$$

Hence, any $\mathbf{V} \in \mathcal{V}_1$ lies in $\mathcal{V}_2$.

Likewise, it can be verified that every $\mathbf{V} \in \mathcal{V}_2$ lies in $\mathcal{V}_1$: For every $\mathbf{V} \in \mathcal{V}_2$ which can be characterized as $\mathbf{V} = \sum_{\ell=1}^{L} \theta_\ell \mathbf{a}_\ell \mathbf{a}_\ell^T$, $\sum_{\ell=1}^{L} \theta_\ell = 1$, set

$$\mathbf{v} = \sum_{\ell=1}^{L} \theta_\ell \mathbf{b}_\ell.$$

It follows from (48) that $\mathrm{Hank}(\mathbf{v}) = \mathbf{V}$. Moreover, this $\mathbf{v}$ satisfies $v_1 = \sum_{\ell=1}^{L} \theta_\ell [\mathbf{b}_\ell]_1 = 1$, and $\mathbf{p}^T \mathbf{v} = \sum_{\ell=1}^{L} \theta_\ell \mathbf{p}^T \mathbf{b}_\ell = 0$.

*E. Proof of Lemma 4*

By Lemma 3, the set $\mathcal{D}$ can be expressed as

$$\mathcal{D} = \left\{ S = \sum_{\ell=1}^{4} \theta_\ell r_\ell \;\middle|\; \sum_{\ell=1}^{4} \theta_\ell \mathbf{a}_\ell \mathbf{a}_\ell^T \succeq \mathbf{0}, \sum_{\ell=1}^{4} \theta_\ell = 1 \right\}$$

This set is a closed convex set, and therefore must be in form of an interval $[L, U]$. The proof is divided into three parts: solving $L$, solving $U$, and integrating the results.

*1) Solving the lower bound:* We find the lower bound $L$ by solving the problem

$$\begin{aligned} L = \min_{\boldsymbol{\theta}} \quad & \sum_{\ell=1}^{4} \theta_\ell r_\ell \\ \mathrm{s.t.} \quad & \sum_{\ell=1}^{4} \theta_\ell \mathbf{a}_\ell \mathbf{a}_\ell^T \succeq \mathbf{0}, \quad \sum_{\ell=1}^{4} \theta = 1. \end{aligned} \tag{49}$$

Let $x_i = \theta_{i+1}$, $i = 1, 2, 3$. Using $\theta_1 = 1 - \sum_{i=1}^{3} x_i$, Problem (49) can be re-expressed as

$$\begin{aligned} L = \min_{\mathbf{x}} \quad & r_1 + \sum_{i=1}^{3} x_i (r_{i+1} - r_1) \\ \mathrm{s.t.} \quad & \mathbf{a}_1 \mathbf{a}_1^T + \sum_{i=1}^{3} x_i (\mathbf{a}_{i+1} \mathbf{a}_{i+1}^T - \mathbf{a}_1 \mathbf{a}_1^T) \succeq \mathbf{0}. \end{aligned} \tag{50}$$

By strong duality, solving (50) is the same as solving its dual which can be shown to be

$$L = \max_{\mathbf{Z} \in \mathbb{S}^3} \ r_1 - \text{tr}(\mathbf{a}_1 \mathbf{a}_1^T \mathbf{Z})$$
$$\text{s.t.} \ \ \mathbf{Z} \succeq \mathbf{0}, \qquad (51)$$
$$\text{tr}[(\mathbf{a}_1 \mathbf{a}_1^T - \mathbf{a}_{i+1}\mathbf{a}_{i+1}^T)\mathbf{Z}] = r_1 - r_{i+1}, \ \ i=1,2,3.$$

From the objective in (51), it is clear that $L = r_1$ if and only if $\mathbf{Z}$ is feasible and satisfies

$$\text{tr}(\mathbf{a}_1 \mathbf{a}_1^T \mathbf{Z}) = 0. \qquad (52)$$

Let us consider the construction of a PSD $\mathbf{Z}$ satisfying (52). Eq. (52) implies that $\mathbf{Z}$ has rank no greater than 2. Thus any such PSD $\mathbf{Z}$ can be represented by

$$\mathbf{Z} = \mathbf{R}\mathbf{R}^T$$

where $\mathbf{R} \in \mathbb{R}^{3 \times 2}$ is such that $\mathbf{R}^T \mathbf{a}_1 = \mathbf{0}$. Such an $\mathbf{R}$ can be parameterized as

$$\mathbf{R} = [\ \mathbf{W}\boldsymbol{\alpha}_1, \ \mathbf{W}\boldsymbol{\alpha}_2\ ],$$

for some $\boldsymbol{\alpha}_1, \boldsymbol{\alpha}_2 \in \mathbb{R}^2$, where

$$\mathbf{W} = \begin{bmatrix} 1 & 1 \\ -1/r_1 & -2/r_1 \\ 0 & 1/r_1^2 \end{bmatrix}.$$

(One can easily check that $\mathbf{a}_1^T \mathbf{W} = \mathbf{0}$, thereby $\mathbf{R}^T \mathbf{a}_1 = \mathbf{0}$.) Therefore, any PSD $\mathbf{Z}$ satisfying (52) can be expressed as

$$\mathbf{Z} = \mathbf{W}\mathbf{G}\mathbf{W}^T \qquad (53)$$

where

$$\mathbf{G} = \begin{bmatrix} a & b \\ b & c \end{bmatrix} = \begin{bmatrix} \boldsymbol{\alpha}_1 & \boldsymbol{\alpha}_2 \end{bmatrix} \begin{bmatrix} \boldsymbol{\alpha}_1^T \\ \boldsymbol{\alpha}_2^T \end{bmatrix} \succeq \mathbf{0}$$

can be any $2 \times 2$ PSD matrix.

By substituting the matrix form in (53) into the equality constraints in (51), we obtain

$$\mathbf{a}_{i+1}^T \mathbf{W}\mathbf{G}\mathbf{W}^T \mathbf{a}_{i+1} = r_{i+1} - r_1, \qquad i = 1,2,3. \qquad (54)$$

We seek to find the sufficient and necessary conditions for satisfying (54). By noticing that

$$\mathbf{W}^T \mathbf{a}_{i+1} = \begin{bmatrix} 1 & -1/r_1 & 0 \\ 1 & -2/r_1 & 1/r_1^2 \end{bmatrix} \begin{bmatrix} 1 \\ r_{i+1} \\ r_{i+1}^2 \end{bmatrix} = \begin{bmatrix} 1 - r_{i+1}/r_1 \\ (1 - r_{i+1}/r_1)^2 \end{bmatrix},$$

Eq. (54) can be decomposed to

$$r_1 - a(r_{i+1}/r_1 - 1) + 2b(r_{i+1}/r_1 - 1)^2 - c(r_{i+1}/r_1 - 1)^3 = 0, \tag{55}$$

for $i = 1, 2, 3$. Let us define a polynomial function

$$f(u) = r_1 - au + 2bu^2 - cu^3.$$

Since the function satisfies $f(r_{i+1}/r_1 - 1) = 0$ for $i = 1, 2, 3$ [cf., Eq. (55)], it permits a factored form

$$f(u) = r_1 \prod_{j=1}^{3} (1 - u/u_j),$$

where

$$u_j = r_{j+1}/r_1 - 1 > 0.$$

By expanding the factored form of $f(u)$ to the polynomial form, we determine (rather tediously) that

$$a = \frac{r_1}{u_1 u_2 u_3}(u_1 u_2 + u_2 u_3 + u_1 u_3) > 0,$$

$$b = \frac{r_1}{2u_1 u_2 u_3}(u_1 + u_2 + u_3) > 0, \quad c = \frac{r_1}{u_1 u_2 u_3} > 0.$$

The remaining part lies in ensuring that the resultant $\mathbf{G}$ is PSD. We already have $a > 0$ and $c > 0$, so the last condition is $b^2 - ac \leq 0$ by Schur complement. With some cumbersome derivations, we show that

$$b^2 - ac = \left(\frac{r_1}{2u_1 u_2 u_3}\right)^2 [u_3 - (\sqrt{u_1} - \sqrt{u_2})^2][u_3 - (\sqrt{u_1} + \sqrt{u_2})^2].$$

In order to achieve $b^2 - ac \leq 0$, we need

$$(\sqrt{u_1} - \sqrt{u_2})^2 \leq u_3 \leq (\sqrt{u_1} + \sqrt{u_2})^2. \tag{56}$$

Summarizing, we have $L = r_1$ if and only if (56) holds.

*2) Solving the upper bound:* The method of the proof is exactly the same as the previous, and hence the detailed derivations are omitted for brevity. Essentially, we consider solving the upper bound

$$U = \max_{\boldsymbol{\theta}} \ \sum_{i=1}^{4} \theta_i r_i$$
$$\text{s.t.} \ \sum_{i=1}^{4} \theta_i \mathbf{a}_i \mathbf{a}_i^T \succeq \mathbf{0}, \ \sum_{i=1}^{4} \theta_i = 1$$

by solving its dual

$$U = \min_{\mathbf{Z} \in \mathbb{S}^3} \ r_4 + \text{tr}(\mathbf{a}_4 \mathbf{a}_4^T \mathbf{Z})$$
$$\text{s.t.} \ \mathbf{Z} \succeq \mathbf{0}, \tag{57}$$
$$\text{tr}[(\mathbf{a}_4 \mathbf{a}_4^T - \mathbf{a}_i \mathbf{a}_i^T) \mathbf{Z}] = r_i - r_4, \ i = 1, 2, 3.$$

From (57) it is shown that $U = r_4$ if and only if

$$(\sqrt{v_2} - \sqrt{v_3})^2 \leq v_1 \leq (\sqrt{v_2} + \sqrt{v_3})^2. \tag{58}$$

where $v_i = 1 - r_i/r_4 > 0$ for $i = 1, 2, 3$.

*3) Combining the conditions:* The final task is to combine the conditions in (56) and (58). We can express (56) as

$$\sqrt{r_3 - r_1} - \sqrt{r_2 - r_1} \leq \sqrt{r_4 - r_1} \leq \sqrt{r_3 - r_1} + \sqrt{r_2 - r_1}.$$

The lower bound is redundant because for any $r_4 > \ldots > r_1 > 0$,

$$\sqrt{r_4 - r_1} \geq \sqrt{r_3 - r_1} \geq \sqrt{r_3 - r_1} - \sqrt{r_2 - r_1}.$$

Moreover, (58) can be expressed as

$$\sqrt{r_4 - r_2} - \sqrt{r_4 - r_3} \leq \sqrt{r_4 - r_1} \leq \sqrt{r_4 - r_2} + \sqrt{r_4 - r_3},$$

and again the lower bound can be shown to be automatically satisfied. We therefore obtain the sufficient and necessary condition in Lemma 4, completing the proof.